\documentstyle[sprocl]{article}   

\input{psfig}

\def\ra{\rightarrow}

\def\al{\alpha}

\def\be{\begin{equation}}
\def\ee{\end{equation}}
\def\bea{\begin{eqnarray}}
\def\eea{\end{eqnarray}}
\newcommand{\bm}{\bibitem}
\newcommand{\om}{\omega}
\renewcommand{\al}{\alpha}
\newcommand{\bt}{\beta}
\newcommand{\lm}{\lambda}
\newcommand{\sg}{\sigma}
\newcommand{\de}{\delta}

\newcommand{\gm}{\gamma}
\newcommand{\ep}{\epsilon}
\newcommand{\Th}{\Theta}

\newcommand{\tq}{\Theta^q}
\newcommand{\tg}{\Theta^g}

\newcommand{\qq}{\bar{q} q}
\newcommand{\hm}{\hat{m}}

\newcommand{\calt}{{\cal T}^{ab}_{\mu\nu}}
    
\newcommand{\vq}{|\vec{q}|}
\newcommand{\la}{\langle}
\renewcommand{\ra}{\rangle}

\newcommand{\OP}{\overline{\langle O \rangle}}
\newcommand{\mrs}{m_{\rho}^2}
\newcommand{\frs}{F_{\rho}^2}

\newcommand{\mn}{4m^2_N}
\newcommand{\qz}{q_{0}}

\begin{document}

\def\thefootnote{\fnsymbol{footnote}} 

\vspace*{-1cm}  
\begin{flushright}
CPT-2000/P.4071
\\
hep-ph/0010329
\\
October 2000 
\end{flushright}

\vspace*{0.6cm}
\title{RHO MESON PROPERTIES IN NUCLEAR MATTER FROM QCD SUM
RULES\footnote{Presented at the Conference on Strong and Electroweak
Matter (SEWM 2000), Marseille, France, 14-17 June 2000. To be
published in the Proceedings. Based on work done in collaboration with 
S.\ Mallik.}}

\author{A.\ NYFFELER}
\address{Centre de Physique Th\'{e}orique, CNRS-Luminy, Case 907 \\
F-13288 Marseille Cedex 9, France \\ 
E-mail: nyffeler@cpt.univ-mrs.fr
}


\maketitle\abstracts{We study the properties of rho mesons in
nuclear matter by means of QCD sum rules at finite density. For
increased sensitivity, we subtract out the vacuum contributions.  With
the spectral function as estimated in the literature, these subtracted
sum rules are found to be not well satisfied. We suppose that Landau
singularities from higher resonance states in the nearby region in
this channel are the cause for this failure.}

\renewcommand{\thefootnote}{\arabic{footnote}}
\setcounter{footnote}{0}

\section{Introduction}

It is believed that the mass and width of particles in a medium differ
from their values in the vacuum. Such an effect might be observable in
various physical systems such as the transition from a quark-gluon
plasma to the hadronic phase in the early universe or in the interior
of neutron stars. Furthermore, such conditions are created in heavy
ion collisions at present and future colliders.

In order to analyze these systems, it is necessary to obtain a good
theoretical understanding of the properties of the particles in a
medium, for instance of the $\rho$-meson. There are various approaches
and models in the literature which predict the behavior of the mass
and width of the $\rho$-meson as function of the temperature and the
density. We will use here the method of QCD sum rules which has been
quite successfully applied in the vacuum case~\cite{SVZ}. This
approach was later extended to finite temperature and
density~\cite{SR_medium}.

Since the medium breaks Lorentz invariance, more operators and unknown
condensates enter on the QCD side of the sum rules (operator product
expansion) and also the hadronic side (spectral function) is only
poorly known in the medium. Nevertheless, one can perhaps gain
insight into the chiral phase transition from experimental data on
$\rho$-mesons in the medium, since these sum rules relate the mass and
width of the $\rho$-meson to condensates like $\langle
\bar q q \rangle$, which serves as an order parameter of chiral
symmetry breaking.
 
Most previous studies~\cite{shift} found at zero temperature a drop of
the $\rho$-meson mass of about 10-20\% 
at nuclear saturation density $\bar n_s$, although in
Ref.~\cite{noshift} no significant shift was observed. In view of
these conflicting results we want to scrutinize here and
in Ref.~\cite{Mallik_Nyffeler} on some aspects of QCD sum rules in the
medium.

\section{Operator product expansion and spectral representation}

We consider the ensemble average of the $T$-product of two vector
currents 
\be \label{VV} 
\calt (q) = i \int{d^4x} e^{iq\cdot x}
\left \langle T\left(V^a_{\mu}(x)
V^b_{\nu}(0)\right) \right \rangle , 
\ee
with $V_\mu^a(x) = \bar{q}(x) \gamma_{\mu}(\tau^a /2) q(x)$. The
ensemble average is defined by $\langle O \rangle = {\rm Tr} [
\exp(-\beta (H-\mu_B N)) O ] / {\rm Tr} [\exp(-\beta (H-\mu_B
N))]$ where $\beta = 1/T$. The nucleon chemical potential has been
denoted by $\mu_B$ and the QCD Hamiltonian and the nucleon number
operator by $H$ and $N$, respectively. The invariant decomposition of
$\calt$ is given by
\be \label{decomposition} 
{\calt} (q) =\de^{ab} (Q_{\mu\nu} T_l + P_{\mu\nu} T_t) \, , 
\ee
with $P_{\mu\nu} = -g_{\mu\nu} + (q_{\mu}q_{\nu} / q^2) - (q^2 /
\bar{q}^2) \tilde{u}_{\mu} \tilde{u}_{\nu}$ and $Q_{\mu\nu} = (q^4 /
\bar{q}^2) \tilde{u}_{\mu}\tilde{u}_{\nu}$, where $\tilde{u}_{\mu} =
u_{\mu} -\om q_{\mu} / q^2$ and $\bar{q} = \sqrt{\om^2 - q^2}$. The
amplitudes $T_{l,t}$ are functions of $q^2$ and $\om = u\cdot q$. We
have introduced the four-velocity of the medium $u_\mu$, in order to
formally restore Lorentz invariance which is broken by the medium.

The sum rules are obtained by equating the short-distance expansion of
the product of the two currents in Eq.~(\ref{VV}), which results in a
series of condensates of QCD operators, with the spectral
representation for the invariant amplitudes $T_{l,t}$, where various
intermediate hadronic states will contribute. 

The short distance expansion of $\calt$ including all operators up to
dimension four can be found in Ref.~\cite{Mallik_Nyffeler}. The
availability of the four-vector $u_\mu$ allows one to construct new
operators as compared to the vacuum case. In addition to the usual
operators $1,\ \qq =\bar{u}u +\bar{d}d, \ G^2 \equiv {\al_s \over
{\pi}} G_{\mu\nu}^a G^{\mu\nu a}$, there are two new operators $\tq
\equiv u^{\mu}\tq_{\mu\nu}u^{\nu}$ and $\tg \equiv
u^{\mu}\tg_{\mu\nu}u^{\nu}$. Here $G_{\mu\nu}^a, \ a=1,...,8$ are the
gluon field strengths and $\al_s =g^2/4\pi$, $g$ being the QCD
coupling constant.  $\Th^{q,g}_{\mu\nu}$ are the quark and the gluon
parts of the traceless stress tensor $\Th_{\mu\nu}^q =
\bar{q}i\gm_{\mu}D_{\nu}q -{\hm\over 4}g_{\mu\nu}\qq$ and 
$\Th_{\mu\nu}^g = -G_{\mu\lm}^c G^{~\lm c}_{\nu} +{1\over 4}g_{\mu\nu}
G_{\al\bt}^c G^{\al\bt c}$, where $\hm$ is the quark mass in the limit
of SU(2) symmetry.  We will take into account the mixing of the
operators $\tq$ and $\tg$ under the renormalization group.

At finite temperature and chemical potential, we can use the Landau
representation~\cite{Landau} of the amplitudes $T_{l,t}$, which is a
spectral decomposition in $q_0$ at fixed $\vq$
\be
T_{l,t}(q_0, \vq) = {1\over{\pi}} \int^{+\infty}_{-\infty}dq^{\prime}_0 
{{\rm Im} T_{l,t}(q^{\prime}_0, \vq)\over{q^{\prime}_0 - q_0 - i\ep}}
{\rm tanh}(\bt \qz^\prime/2) , 
\ee
up to subtraction terms. There will be contributions in the integral
from various intermediate hadronic states, first of all from the
$\rho$-meson. The effects of the medium can be parametrized by
employing the operator relation
\be
V^a_{\mu}(x) = m_{\rho}^{\star}F_{\rho}^{\star}\rho^a_{\mu}(x) , 
\ee
where $m_{\rho}^{\star}$ and $F_{\rho}^{\star}$ denote the in-medium
mass and width of the $\rho$-meson. This will generate the usual
$\delta$-function contribution in ${\rm Im} T_{l,t}$. 

Moreover, in the nuclear medium there will be a contribution from $N
\bar{N}$ intermediate states. In the vacuum this contribution
is small, coming from the cut beginning at threshold, $q_0^2=4 m_N^2
+\vq^2$. However, in the nuclear medium the currents can also interact
with real nucleons to give rise to a short cut around the origin,
$-\vq< q_0 <+\vq$. The evaluation of this contribution can be found in
Ref.~\cite{Mallik_Nyffeler}. 

Following Refs.~\cite{shift,noshift} we ignore the states
$N\bar{N}^\star$ with resonances $N^\star$. Furthermore, below we will
restrict ourselves to sum rules at finite nuclear density but at zero
temperature, therefore the $2\pi$ contribution will be eliminated
after subtracting out the corresponding vacuum sum rules.

\section{Sum rules} 

We obtain the sum rules by equating the spectral representation and
the operator product expansion for the two invariant amplitudes
$T_{l,t}$. As usually done in the literature~\cite{SVZ}, we take the
Borel transform of both sides in order to enhance the contribution
from the lowest lying resonance, here the $\rho$-meson, and to
suppress the contributions from higher dimensional operators. In
general, this can only be achieved within a certain region for the so
called Borel parameter $M$. In contrast to earlier
work~\cite{shift,noshift} we subtract the vacuum sum rules, assuming
that the contribution from the QCD continuum will practically drop out
in this way. For $T \to 0$ and $\mu_B > 0$ the $2\pi$ contribution
will also cancel and we obtain the following sum rules in the
limit $\vq \rightarrow 0$ where the expressions simplify considerably:
\bea
\overline{F^{\star 2}_\rho e^{-{m_\rho^{\star 2} \over M^2}}} 
+ {1\over {24 \pi^2}} \int_{\mn}^{4\mu_B^2} ds (1-e^{-{s\over M^2}})
\sqrt{1- {\mn \over s}}(1+ {2m_N^2 \over s}) & = & {\OP\over M^2},
\label{SR1} \\
\overline{m_{\rho}^{\star 2}F^{\star 2}_\rho e^{-{m_\rho^{\star
2} \over M^2}}} 
- {1\over {24 \pi^2}} \int_{\mn}^{4\mu_B^2} ds \, s \,  
e^{-{s\over M^2}} \sqrt{1-{\mn \over s}}(1+ {2m_N^2 \over s}) 
&=&-\OP, \label{SR2} 
\eea
with 
\be
\la O \ra = {1\over 2} \hat{m} \la \qq \ra + {\la G^2 \ra \over
24}+ {2\over 11} \left( \la \Th \ra + \lm(M^2)  \la {8\over 3} \Th^q -\Th^g
\ra \right), 
\ee
where $\overline{\la O \ra} = \la O \ra -\la 0|O|0\ra$ and $\Th =
\Th^q + \Th^g$. The mixing under the renormalization group is taken
into account by $\lm(M^2) = ( \al_s(\mu^2) / \al_s(M^2))^{-d/2b}$ with
$d={4\over 3}({16\over 3} + n_f)$ and $b =11- {2\over 3}n_f$. We will take
$\mu = 1~{\rm GeV}$ below.  

In the linear density approximation, we expand the mass and width
and the condensates up to first order in the nucleon number density
$\bar n$,
\be \label{def_abC} 
m^{\star}_{\rho} = m_{\rho} (1+ a {\bar{n}\over \bar{n}_s}), \quad
F^{\star}_{\rho} = F_{\rho} (1+ b {\bar{n}\over \bar{n}_s}), 
\quad
\OP = C\bar{n} \, , 
\ee 
where $\bar{n}_s = (110~{\rm MeV})^3$ denotes the nuclear saturation
density. The coefficient $C$ in Eq.~(\ref{def_abC}) is given by 
\be
C = {\sg\over 2}-{1\over 27}(m_N -\sg) +{3\over 22}m_N
\left( A^q + A^g + \lm(M^2) ({8\over 3} A^q - A^g)  \right) , 
\ee
where $\sigma = \la p | \hm (\bar u u + \bar d d) | p \ra / (2 m_N)
\simeq 45 \ {\rm MeV}$ denotes the sigma term~\cite{Gasser}. The
constants $A^{q,g}$ are defined by $\la p|\Th_{\mu\nu}^{q,g} |p\ra =2
A^{q,g} (p_{\mu} p_{\nu} -{1\over 4} g_{\mu\nu} p^2)$. From parton
distribution functions~\cite{MRST} one obtains the values $A^q = 0.62$
and $A^g = 0.35$.

From the sum rules~(\ref{SR1}) and (\ref{SR2}) we finally obtain the
expressions
\bea
a &= &-{\bar{n}_s \over 2 \frs} e^{{\mrs\over M^2}} \! \left[ C ({1\over
\mrs}+{1\over M^2})  
-{1\over{4m_N}} - ({m_N\over \mrs} - {1\over {4m_N}})
e^{-{\mn\over M^2}}\right]\!\! ,  
\label{SRa} \\ 
b &= & - {\bar{n}_s \over 2 \frs} e^{{\mrs\over M^2}} \! \left[ C
{\mrs\over M^4} + {1\over {4m_N}} (1-{\mrs\over M^2}) 
- \left({m_N\over {M^2}} + {1\over {4m_N}} (1-{\mrs\over M^2})
\right) \! e^{-{\mn\over M^2}}
\right]\!\! .\nonumber \\
&&\label{SRb}
\eea

%\vspace*{-0.2cm}
\section{Discussion and Conclusions}

\begin{figure}[!h]
\centerline{\psfig{figure=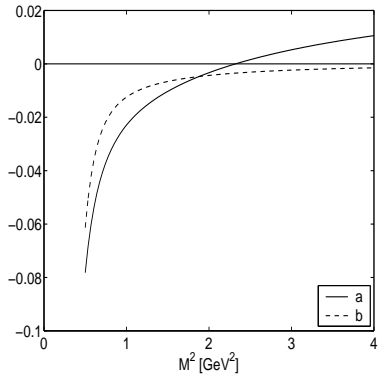,height=5cm,width=5cm}}
\caption{Coefficients of $\bar{n} / {\bar{n}_s}$ in the expansion of 
$m_{\rho}^{\star}$ and $F_{\rho}^{\star}$ from Eq.~(\ref{def_abC}) as
function of the Borel parameter $M$.} \label{fig_ab}
\end{figure}

In Fig.~\ref{fig_ab} we have plotted the coefficients $a$ and $b$ as 
function of the Borel parameter $M.$
We note that there is no sign of constancy of $a$ and $b$ in any
region of $M.$ We thus conclude that the sum rules in Eqs.~(\ref{SR1})
and (\ref{SR2}) cannot give any reliable information about the density
dependence of $m_\rho$ and $F_\rho$.

We have included only operators up to dimension four in the sum rules
above, but taken into account their mixing which is numerically
relevant for $M^2 \neq \mu^2$.  Contributions from higher dimensional
operators should, however, be relatively small for $M^2 > 1~{\rm
GeV}^2$. A more detailed analysis is in
progress~\cite{Mallik_Nyffeler}.

Therefore, the failure of our subtracted sum rules can presumably be
traced to the hadronic spectral side which is not adequately
saturated.  In fact, higher resonance states $N \bar N^\star$ will
also contribute a cut for $q^2 \leq (m_N^\star - m_N)^2$. The problem
with the inclusion of such contributions is, however, that more
unknown couplings and masses will enter in the medium.

The sum rules at finite density are given by the vacuum sum rules
perturbed by small terms proportional to the density.  Since the
vacuum sum rules are stable~\cite{SVZ}, this guarantees the stability
of the sum rules at finite density as observed in previous
work~\cite{shift,noshift}. In contrast, the subtracted sum rules above
are much more sensitive to errors in or omissions of any terms.

\vspace*{-0.25cm}
\section*{Acknowledgments}
This article is based on joint work with S.\ Mallik. I would like to
thank H.\ Leutwyler and P.\ Minkowski for helpful discussions. This
work was supported in part by Schweizerischer Nationalfonds.

\vspace*{-0.25cm}
\section*{References}

\end{document}